\documentclass[11pt,notitlepage,longbibliography,nofootinbib,prd]{revtex4-1}
\usepackage{verbatim}
\usepackage{amsmath,amssymb,amsfonts}
\usepackage{graphicx}
\usepackage{color}
\usepackage[colorlinks=true,linkcolor=black,urlcolor=blue,citecolor=blue]{hyperref}

\def\mf{m_{\rm f}}
\def\mc{\mathcal}
\def\del{\partial}

\newcommand{\half}{\frac{1}{2}}

\def\a{\alpha}
\def\b{\beta}
\def\d{\delta}
\def\D{\Delta}
\def\e{\epsilon}

\def\l{\lambda}
\def\L{\Lambda}
\def\m{\mu}
\def\n{\nu}

\def\O{\Omega}

\def\P{\Phi}
\def\vp{\varphi}
\def\s{\sigma}

\def\th{\theta}

\def\coeff#1#2{{\textstyle {\frac {#1}{#2}}}}
\def\GN{G}

\begin{document}

\title{\Large{Einstein's Equations in Matter}}

\author{Pavel Kovtun}
\email{pkovtun@uvic.ca}
\author{Ashish Shukla}
\email{ashish@uvic.ca}
\affiliation{Department of Physics \& Astronomy,  University of Victoria, 3800 Finnerty Road, Victoria, BC,  V8P 5C2, Canada}

\begin{abstract}
\noindent 
Einstein's equations in matter are gravitational analogues of Maxwell's equations in matter, providing an effective classical description of gravitational fields. We derive Einstein's equations in matter for relativistic fluids, and use them to illustrate how the Tolman-Oppenheimer-Volkoff equations are modified by the matter's response to curvature. For a gas of massive fermions, we evaluate how the effective Newton's constant and other susceptibilities depend on the temperature and density. In anti-de Sitter space, we study the $O(1/(T\ell)^2)$ corrections to the geometries sourced by perfect fluids, and illustrate the breakdown of hydrostatics in AdS at small temperatures.
\end{abstract}

\maketitle

\section{Introduction}
\label{Intro}
Matter affects the dynamics of both electromagnetic and gravitational fields. For electromagnetism, the effective classical description in many materials is provided by the so-called Maxwell's equations in matter \cite{Jackson:1998nia}. In the effective description, the microscopic Maxwell's equations $\nabla_{\!\nu} F^{\mu\nu} = J^\mu$ are replaced by $\nabla_{\!\nu} H^{\mu\nu} = J^\mu_{\rm free}$, where the tensor $H^{\mu\nu}$ differs from $F^{\mu\nu}$ by substituting the electric field ${\bf E} \to {\bf D}$, and the magnetic field ${\bf B}\to {\bf H}$. The current $J^\mu_{\rm free}$ is the current of ``free'' charges, and the vectors ${\bf D}$, ${\bf H}$ are to be expressed in terms of ${\bf E}$, ${\bf B}$ through the so-called constitutive relations that are specific to a given material. The constitutive relations reflect the existence of both electric and magnetic polarization. As a simple example, consider the Maxwell action in flat space, $S_{\rm eff} = -\coeff14 \int d^4x \, F_{\mu\nu} F^{\mu\nu}$, or in terms of the electric and magnetic fields
\begin{equation}
  S_{\rm eff} = \half \int\!\!d^4x \left( {\bf E}^2 - {\bf B}^2 \right).
\end{equation}
In the presence of matter, boost invariance is broken by the matter rest frame, and a more general action is allowed by rotation invariance:
\begin{equation}
\label{eq:emaction-2}
  S_{\rm eff} = \half \int\!\!d^4x \left(\varepsilon_{\text e} {\bf E}^2 - \frac{{\bf B}^2}{\mu_{\text m}} \right).
\end{equation}
The coefficients $\varepsilon_{\text e}$ (electric permittivity) and $\mu_{\text m}$ (magnetic permeability) can in principle be calculated from the fundamental microscopic theory. The action gives rise to the constitutive relations ${\bf D} = \varepsilon_{\text e} {\bf E}$ and ${\bf B} = \mu_{\text m} {\bf H}$ (see e.g.~\cite{Kovtun:2016lfw} for a covariant discussion). This leads to the refractive index $\sqrt{\varepsilon_{\text e} \mu_{\text m}}$ for electromagnetic waves propagating through matter. 

One can ask similar questions about gravity. While the naive notions of electric and magnetic  polarizations do not have a gravitational analogue, one can ask about the effective description of gravitational fields in matter. Just like in electromagnetism the definition of the macroscopic tensor $H^{\mu\nu}$ boils down to the structure of the macroscopic current~$J^\mu$, one can ask about the structure of the macroscopic energy-momentum tensor $T^{\mu\nu}$ in the Einstein's equations $G_{\mu\nu} = \kappa^2\, T_{\mu\nu}$, where $G_{\mu\nu}$ is the Einstein tensor, and $\kappa^2 \equiv 8\pi\GN$, with $\GN$ the Newton's constant. As an example, consider the following effective action for equilibrium gravity-matter configurations,
\begin{equation}
\label{eq:S1}
  S_{\rm eff} = \int \!\!d^{d+1}x\,\sqrt{-g} \left[ \frac{1}{2\kappa^2} R + p(T) \right],
\end{equation}
where $R$ is the Ricci scalar, and $p(T)$ is the pressure as a function of temperature, given by the equation of state. As we will show shortly, this action gives rise to Einstein's equations sourced by the perfect fluid energy-momentum tensor, $G^{\mu\nu} = \kappa^2\left[(\epsilon{+}p) u^\mu u^\nu + p g^{\mu\nu} \right]$, where $\epsilon \equiv -p + T \partial p/\partial T$ is the energy density, derived from the pressure by the standard Euler relation of thermodynamics. However, in curved space, there is more to thermodynamics than just the equation of state. The perfect fluid model of matter ignores the fact that the energy-momentum tensor of macroscopic matter in general depends on curvature, which is true even for an ideal (quantum) gas of spinless particles. As an example, note that the action~(\ref{eq:S1}) is only an effective macroscopic action, and the coefficient in front of $R$ is not required to be constant due to the presence of matter. More generally, one can have
\begin{equation}
\label{eq:S2}
  S_{\rm eff} = \int \!\!d^{d+1}x\,\sqrt{-g} \left[ f(T) R + p(T) \right]\,,
\end{equation}
where $f(T)$ plays the role of a gravitational susceptibility. Just like the pressure $p(T)$, the susceptibility coefficient $f(T)$ can in principle be evaluated from the fundamental microscopic theory, and will depend on temperature. The vacuum contributions to $p(T)$ and $f(T)$ have to be matched to the cosmological constant and the Newton's constant respectively, $p(T) = (-2\Lambda)/2\kappa^2 + p_{\rm matter}(T)$, $f(T)=1/2\kappa^2 + f_{\rm matter}(T)$. As the temperature $T$ depends on the metric,%
\footnote{
For matter in equilibrium, one has $T=T_0/\sqrt{-g_{00}}$ in suitable coordinates, see e.g.~\cite{Landau:1980mil}. 
}
the Einstein's equations obtained by varying (\ref{eq:S2}) with respect to the metric will be sourced by the energy-momentum tensor which is not of the perfect fluid form, due to $f_{\rm matter}(T)$.

In what follows, we will make the above intuition precise. Focusing on the simplest form of relativistic matter which is locally rotation-invariant (that is, a fluid), we will derive the Einstein equations which govern the effective gravitational fields. Among other things, we will explicitly evaluate $f_{\rm matter}(T)$ for an ideal gas of massive fermions at non-zero temperature and density. We will then consider an application of the formalism to spherical equilibria of gravitating matter. [Convention: We work in natural units $c = \hbar = 1$. The metric signature is mostly plus.]

\section{Einstein's equations in matter}
\label{EMEquations}
We consider a macroscopic system that has degrees of freedom which couple to the metric $g_{\mu\nu}$ and to an Abelian gauge field $A_\mu$. Our procedure for deriving the effective Einstein-Maxwell equations will be as follows. (I) We first consider the matter in thermal equilibrium, subject to external, time-independent $g_{\mu\nu}$ and $A_\mu$. The system is then characterized by the grand canonical partition function $Z$, and the generating functional $W[g,A]=-i\ln Z$. The generating functional (which is the grand canonical potential, up to a factor of $T$) is extensive in the thermodynamic limit, and can be written as an integral of a local density. (II)~We identify the equilibrium generating functional with the equilibrium effective action, $W[g,A] = S_{\rm eff}[g,A]$. With the fields $g$ and $A$ now dynamical, the equations of motion which follow from the equilibrium effective action will describe equilibrium configurations of the metric and the gauge field. (III) We then argue that for time dependent processes that are sufficiently slow, the equations of motion that describe equilibrium configurations may serve as a useful approximation to describe near-equilibrium configurations. The resulting Einstein-Maxwell equations will contain perfect fluid dynamics, electromagnetic polarization, and the gravitational susceptibilities that parametrize the response of the matter's pressure to curvature. The resulting equations clearly do not include any dissipative processes; the latter, parametrized by the dissipative transport coefficients such as the electrical conductivity and the viscosity, can be introduced in the constitutive relations {\it a~posteriori}. Step (III) is just the standard construction of fluid dynamics as an extension of local equilibrium thermodynamics. If we choose to keep only $g$ (but not $A$) dynamical, we have the Einstein (rather than Einstein-Maxwell) equations in matter, with $A$ as a background field which can couple for example to the baryon number current.

\subsection{The generating functional}
\label{EqGenFunc}
Consider matter in thermal equilibrium, described by the generating functional $W[g,A]$. The generating functional encodes equilibrium (zero-frequency) correlation functions of the energy-momentum tensor and the conserved $U(1)$ current which couple to the external metric $g_{\m\n}$ and the external gauge field $A_\m$. In the context of relativistic hydrodynamics such equilibrium generating functionals were first discussed in \cite{Banerjee:2012iz, Jensen:2012jh}; the present discussion makes use of the setup in \cite{Jensen:2012jh, Jensen:2013kka, Hernandez:2017mch, Kovtun:2018dvd}. 

Thermal equilibrium is characterized by a timelike Killing vector $V^\m$, which in suitable coordinates (local rest frame of the fluid) takes the form $V^\m = (1,{\bf 0})$. The matter velocity $u^\mu$, temperature $T$, and the chemical potential $\mu$ are defined via
\begin{equation}
\label{eq:uTmu}
  u^\mu = \frac{V^\m}{\sqrt{-V^2}}\,,\
  T = \frac{1}{\b_0 \sqrt{-V^2}}\,,\ 
  \m = \frac{ V^\rho A_\rho + \L_V}{\sqrt{-V^2}}\,,
\end{equation}
where the constant $\beta_0$ sets the normalization of temperature, and $\Lambda_V$ is a gauge function introduced to ensure the gauge invariance of the chemical potential. The fact that the system is in equilibrium is captured by the conditions
\begin{equation}
\label{eq:equil-cond}
  L_V g_{\m\n}=0\,,\quad L_V A_\m + \partial_\m \L_V = 0\,,
\end{equation}
where $L_V$ denotes the Lie derivative with respect to the timelike vector $V^\m$. 

The generating functional is extensive in the thermodynamic limit, and can therefore be expressed as an integral of a local density,
\begin{equation}
\label{eq:W1}
  W[g,A] = \int\!\!d^{d+1}x\,\sqrt{-g}\; {\cal F}[g,A]\,,
\end{equation}
where $\mc{F}[g,A]$ is a local function of the external sources $g_{\m\n}, A_\m$. The variation of the generating functional with respect to the sources defines the energy-momentum tensor and the conserved current of the fluid,
\begin{equation}
\label{defTJ}
T^{\m\n} = \frac{2}{\sqrt{-g}} \, \frac{\d W}{\d g_{\m\n}}\, , \quad J^\m = \frac{1}{\sqrt{-g}} \, \frac{\d W}{\d A_{\m}}.
\end{equation}
Provided that the system of interest does not have any anomalies, the generating functional is both diffeomorphism and gauge invariant. As a consequence of this we have the conservation laws
\begin{subequations}
\label{eq:TJ-cons-1}
\begin{align}
  & \nabla_{\!\m} T^{\m\n} = F^{\n\l}J_\l\,,\\
\label{eq:J-cons-1}
  & \nabla_{\!\m} J^\m = 0\,,
\end{align}
\end{subequations}
with $F_{\m\n} \equiv \del_\m A_\n - \del_\n A_\m$. 

When the sources $g_{\m\n}, A_\m$ vary on length scales much longer than the correlation length of the system, the density $\mc{F}[g,A]$ admits an expansion in terms of the derivatives of the external sources. Thus, finding the generating functional up to a given order in derivatives boils down to finding the diffeomorphism- and gauge-invariant objects made out of the sources $g_{\m\n}, A_\m$, as well as the hydrodynamic variables of eq.\ \eqref{eq:uTmu}. 

Being in thermal equilibrium implies certain constraints on the hydrodynamic variables, such as
\begin{equation}
u^\l \del_\l T = 0, \quad u^\l \del_\l \m = 0,
\end{equation}
i.e.~the temperature and the chemical potential are time independent. The expansion $\nabla{\cdot}u$ and the shear tensor $\s^{\m\n} = \D^{\m\a} \D^{\n\b} (\nabla_\a u_\b + \nabla_\b u_\a - \coeff23 \D_{\a\b} \nabla{\cdot} u)$ both vanish in equilibrium, ensuring that there is no entropy production due to the bulk or shear viscosities in equilibrium. The spatial projector is $\Delta^{\mu\nu} \equiv g^{\mu\nu} + u^\mu u^\nu$.

For the derivative expansion of the generating functional, we follow the notation of~\cite{Kovtun:2018dvd}. In four spacetime dimensions, and up to second order in derivatives of the sources, the equilibrium generating functional takes the form
\begin{equation}
\label{eq:W2}
  W[g,A] = \int\!\!d^{4}x\,\sqrt{-g}\left[ p + \sum_{n=1}^9 f_n\, s_n^{(2)}\right] + \cdots
\end{equation}
where the dots denote terms that are cubic or higher order in derivatives. The pressure $p=p(T,\m)$ is a function of temperature and chemical potential, given by the equation of state. There are no first-order invariants. The $s_n^{(2)}$ denote the nine independent second-order invariants appearing in the generating functional \cite{Banerjee:2012iz}, and the $f_n = f_n(T,\mu)$ denote the corresponding thermodynamic susceptibilities. Table \ref{table:2ndOrder} summarizes the nine invariants, along with their transformation properties under parity $\mc{P}$, charge conjugation $\mc{C}$, and time reversal $\mc{T}$. The notation in the table is as follows: the acceleration is $a^\m \equiv u^\n \nabla_\n u^\m$, the electric field is $E_\mu = F_{\mu\nu} u^\nu$, the magnetic field is $B^\m \equiv \coeff12 \e^{\m\n\a\b} u_\n F_{\a\b}$, and the vorticity vector is $\O^\m \equiv \e^{\m\n\a\b} u_\n \nabla_{\!\a} u_\b$. The pressure and the susceptibilities have to be determined from the underlying microscopic theory. See \cite{Romatschke:2009ng, Moore:2010bu, Moore:2012tc, Megias:2014mba, Buzzegoli:2017cqy, Kovtun:2018dvd, Shukla:2019shf} where some of these susceptibilities have been computed for free quantum fields. 

\begin{table}
\begin{center}
\def\arraystretch{1.2}
\setlength\tabcolsep{4pt}
\begin{tabular}{|c|c|c|c|c|c|c|c|c|c|}
 \hline
 \hline
 $n$ & 1 & 2 & 3 & 4 & 5 & 6 & 7 & 8 & 9\\ 
 \hline
 \hline
 $s^{(2)}_n$
 & $R$   %
 & $a^2$  %
 & $\Omega^2$
 & $B^2$ 
 & $B{\cdot}\Omega$
 & $E^2$
 & $E{\cdot}a$
 & $B{\cdot}E$
 & $B{\cdot}a$
 \\
 \hline
  $\mc{P}$  & $+$ & $+$ & $+$ & $+$ & $+$ & $+$ & $+$ & $-$ & $-$\\
  \hline
  $\mc{C}$  & $+$ & $+$ & $+$ & $+$ & $-$ & $+$ & $-$ & $+$ & $-$\\
  \hline
  $\mc{T}$  & $+$ & $+$ & $+$ & $+$ & $+$ & $+$ & $+$ & $-$ & $-$\\
  \hline
\end{tabular}
\end{center}
\caption{The nine second-order invariants appearing in the derivative expansion eq.\ \eqref{eq:W2}. $\mc{P}, \mc{C}, \mc{T}$ respectively denote the eigenvalues under parity, charge conjugation and time reversal of the corresponding invariant.}
\label{table:2ndOrder}
\end{table}

\subsection{The energy-momentum tensor}
We can now compute the energy-momentum tensor following from the variation of the generating functional eq.\ \eqref{eq:W2}. Isolating the Einstein tensor, we have
\begin{equation}
\label{eq:Tmatter-def}
  T^{\mu\nu} = -\frac{1}{\kappa_{\rm eff}^2} \left( R^{\mu\nu} - \coeff12 R g^{\mu\nu} \right) + T_{\rm m}^{\mu\nu}\,,
\end{equation}
which defines the ``matter'' contribution $T_{\rm m}^{\mu\nu}$, and $1/\kappa_{\rm eff}^2 \equiv 2f_1(T,\mu)$. We can decompose $T_{\rm m}^{\mu\nu}$ with respect to the fluid velocity $u^\mu$ as
\begin{equation}
\label{eq:Tmn}
  T_{\rm m}^{\m\n} = {\cal E}_{\rm m} u^\m u^\n + {\cal P}_{\rm m}\D^{\m\n} +{\cal Q}_{\rm m}^\m u^\n + {\cal Q}_{\rm m}^\n u^\m + {\cal T}_{\rm m}^{\m\n},
\end{equation}
where the ``matter'' energy density is ${\cal E}_{\rm m}\equiv u_\mu T_{\rm m}^{\mu\nu}u_\nu$, the pressure is ${\cal P}_{\rm m}\equiv \frac13 \Delta_{\mu\nu}T_{\rm m}^{\mu\nu}$, the energy flux ${\cal Q}_{\rm m}^\mu \equiv - \Delta^\mu_{\alpha} T_{\rm m}^{\alpha\beta}u_\beta$ is transverse to $u^\mu$, and the stress ${\cal T}_{\rm m}^{\mu\nu} \equiv T_{\rm m}^{\langle \mu\nu\rangle}$ is transverse to $u_\mu$, symmetric, and traceless. The brackets denote the symmetric transverse traceless part, $X_{\langle \mu\nu\rangle} \equiv \frac12 (\Delta_{\mu\alpha} \Delta_{\nu\beta} + \Delta_{\nu\alpha} \Delta_{\mu\beta} -\frac23 \Delta_{\mu\nu} \Delta_{\alpha\beta}) X^{\alpha\beta}$.

For matter that is not coupled to external electric or magnetic fields, the only relevant two-derivative invariants are $f_{1,2,3}$. The matter energy-momentum tensor eq.\ \eqref{eq:Tmn} coming from the variation of the generating functional eq.\ \eqref{eq:W2} then has the form
\begin{subequations}
\label{eq:cr-statics}
\begin{align}
&\mc{E}_{\rm m} = \e + f_1' R + \big( 4f_1' + 2f_1'' - f_2 - f_2'\big) a^2 +\big( f_1'-f_2-3f_3+f_3'\big) \O^2- 2 \left(f_1'-f_2\right) u^\a R_{\a\b} u^\b , \label{stressel}\\
&\mc{P}_{\rm m} = p - \frac{1}{3} \left(2f_1' + 4 f_1'' - f_2 \right) a^2 - \frac{1}{3} (2f_1'+f_3) \O^2 + \frac{4}{3} f_1' u^\a R_{\a\b} u^\b , \label{stressP}\\
&\mc{Q}_{{\rm m}\,\m} = \big(f_1'+2f_3'\big) \e_{\m\n\a\b} a^\n u^\a \O^\b + 4f_3 \D^\n_\m R_{\n\s} u^\s, \label{stressQ}\\
&\mc{T}_{{\rm m}\,\m\n} = \left( 4 f_1' + 2 f_1'' - 2 f_2\right) a_{\langle \m} a_{\n \rangle} - \frac{1}{2} \big( f_1' - 4f_3\big) \O_{\langle\m} \O_{\n\rangle} + 2 f_1' u^\a R_{\a\langle \m \n \rangle \b} u^\b, \label{stressT}
\end{align}
\end{subequations}
with
\begin{equation}
\begin{split}
\label{defprime}
&f_n' = T \, \frac{\del f_n}{\del T} + \m \, \frac{\del f_n}{\del \m},\\
&f_n'' =  T^2 \, \frac{\del^2 f_n}{\del T^2}+ 2\m T \,\frac{\del^2 f_n}{\del \m \, \del T}  + \m^2 \, \frac{\del^2 f_n}{\del \m^2},
\end{split}
\end{equation}
and $\e = - p + T\, \del p/\del T + \m \, \del p/\del \m$ is the zeroth-order energy density.

\subsection{The effective Einstein's equations}
We now promote the equilibrium generating functional to be the equilibrium effective action for the metric, $W[g,A] = S_{\rm eff}[g,A]$. This makes the metric a dynamical variable whose equations of motion are determined by the variational principle $\delta_g S_{\rm eff}[g,A] = 0$, or simply $T^{\m\n}[g,A] = 0$. According to the definition of the ``matter'' contribution in eq.~(\ref{eq:Tmatter-def}), the equations of motion are
\begin{equation}
\label{eq:EE-1}
  R^{\mu\nu} - \coeff12 R g^{\mu\nu}  = \kappa_{\rm eff}^2 T_{\rm m}^{\mu\nu}\,,
\end{equation}
where $T^{\mu\nu}_{\rm m}$ contains the perfect fluid, as well as the curvature-dependent contributions. The coefficient $\kappa_{\rm eff}^2 = 1/(2f_1)$ should be interpreted as ($8\pi$ times) the effective Newton's constant which now depends on temperature and chemical potential due to the presence of matter. For a fluid not subject to external electric and magnetic fields, $T^{\mu\nu}_{\rm m}$ is given by eqs.~(\ref{eq:Tmn}), (\ref{eq:cr-statics}). 

Equations \eqref{eq:EE-1} are the effective Einstein's equations in equilibrium matter, analogous to the Maxwell's equations in matter for electro- and magneto-statics. The susceptibilities $f_{1,2,3}(T,\mu)$ characterize the leading-order (in derivatives) equilibrium response of the energy-momentum tensor to curvature. We emphasize that the effective Einstein's equations \eqref{eq:EE-1} do not arise from any modification of Einstein's gravity, but are simply a reflection of the gravitational properties of normal matter. 

We do not write down the three-derivative and higher-order terms in $T^{\mu\nu}_{\rm m}$ which would come with their own susceptibilities, as in the standard effective field theory. The non-equilibrium terms in the constitutive relations of $T^{\mu\nu}_{\rm m}$ do not follow from the equilibrium generating functional, and have to be added separately, as is done in standard relativistic hydrodynamics.

In practice, one expects the effect of the susceptibilities on the gravitational dynamics to be very small. Indeed,  $p = (-2\Lambda)/2\kappa^2 + p_{\rm m}(T,\mu)$, $f_1=\frac12 M_{\rm Pl}^2 + f_{1,{\rm m}}(T,\mu)$, where $M_{\rm Pl} =  1/\sqrt{8\pi G}$ is the Planck mass, and the subscript ``m'' denotes the matter contribution. For non-cosmological applications one usually sets $\Lambda=0$, while the  temperatures and densities are such that $T,\mu \ll M_{\rm Pl}$. On the other hand, the susceptibilities are more relevant for non-gravitational dynamics of relativistic fluids where $f_{1,2,3}$ show up as thermodynamic transport coefficients, see e.g.~\cite{Kovtun:2018dvd}. When studying matter in Anti-de Sitter space, one has to keep $\Lambda$ non-zero, and the gravitational susceptibilities lead to interesting phenomena, as we discuss below.

Let us now discuss the application of the above theory to non-rotating spherically symmetric equilibrium states of gravitating matter.

\section{Spherical equilibria}

We are interested in equilibrium states of gravitating matter. The simplest way to find such equilibria in classical general relativity is by solving the Einstein's equations
\begin{equation}
  G_{\mu\nu} = \kappa^2 T_{\mu\nu}\,,
\end{equation}
given a certain equation of state that determines the energy-momentum tensor $T^{\mu\nu}$ on the right-hand side. The problem has been explored extensively, starting with the treatment by Tolman many years ago~\cite{Tolman}.
Spherical equilibrium for a perfect fluid with the equation of state given by free fermions at zero temperature was studied by Oppenheimer and Volkoff~\cite{Oppenheimer:1939ne}, and is by now textbook material~\cite{Shapiro:1983du}. Let us briefly review the setup.

The most general rotationally invariant static metric can be written as
\begin{equation}
\label{eq:ss1}
  ds^2 = -A(r) dt^2 + B(r) dr^2 + r^2 d\O^2\,,
\end{equation}
where $d\O^2 = d\th^2 + \sin^2\th \, d\varphi^2$ is the line element on the unit two-sphere. The metric is assumed to be sourced by the perfect fluid energy-momentum tensor. In the above static coordinates, $T_{\mu}^{\ \, \nu} = {\rm diag} (-\epsilon, p, p, p)$, where the energy density $\epsilon$ and the pressure $p$ are functions of $r$. 
The metric eq.\ (\ref{eq:ss1}) can be written in terms of the ``potential'' $\P(r)$ and the ``mass function'' $m(r)$ as 
\begin{equation}
\label{eq:ss2}
  ds^2 = -e^{2\Phi(r)} dt^2 + \left(1-\frac{2\GN m(r)}{r} \right)^{-1} dr^2 + r^2 d\Omega^2.
\end{equation}
The $tt$ and $rr$ components of the Einstein's equations give, respectively
\begin{align}
\label{eq:eett}
  & \frac{dm}{dr} = 4\pi r^2 \epsilon\,,\\
\label{eq:eerr}
  & \frac{d\P}{dr} = \frac{\GN m}{r^2} \frac{1+ 4\pi r^3 p/m}{1- 2\GN m/r}\,.
\end{align}
Due to the spherical symmetry of the problem, the $\th\th$ and $\vp\vp$ components of the Einstein's equations carry the same information. The $\th\th$ equation contains $d^2\P/dr^2$, which can be eliminated by combining it with the $r$ derivative of eq.\eqref{eq:eerr}. This gives the equation
\begin{align}
\label{eq:eethth}
  \frac{dp}{dr} = -\frac{\GN m}{r^2} \frac{\epsilon + p}{1- 2\GN m/r} \left(1+\frac{4\pi r^3 p}{m} \right).
\end{align}

We thus have three equations (\ref{eq:eett}), (\ref{eq:eerr}) and (\ref{eq:eethth}) for four functions $m(r)$, $\Phi(r)$, $\epsilon(r)$, and $p(r)$, which are the Tolman-Oppenheimer-Volkoff (TOV) equations. The conservation of the energy-momentum tensor is automatic thanks to $\nabla_{\!\mu} G^{\mu\nu} = 0$, and does not have to be imposed as an independent condition. The extra equation must therefore come from the equation of state.

If we can express $\epsilon=\epsilon(p)$ from the equation of state, then eqs.\ (\ref{eq:eett}) and (\ref{eq:eethth}) give two coupled equations for the two unknown functions $m(r)$ and $p(r)$. These can be solved for $m(r), p(r)$ with appropriate boundary conditions, and the resulting solutions inserted into eq.\ \eqref{eq:eerr} to determine the potential $\P(r)$.

However, the equation of state can not usually be written in the form $\e=\e(p)$. Namely, in the grand canonical ensemble we have $\epsilon=\epsilon(T,\mu)$, $p=p(T,\mu)$, where $T$ is the temperature and $\mu$ is the chemical potential (assuming one species of particles). In other words, $p$ is a function of two independent variables, while $\epsilon$ is just one combination of the two variables. So we have three equations, and four unknown functions $T,\mu,m,\Phi$. To address this issue, note that eqs.\ (\ref{eq:eerr}) and (\ref{eq:eethth}) imply $dp/dr=-(\epsilon+p)\,d\P/dr$. Extensivity in the grand canonical ensemble implies $\epsilon+p=T\frac{\partial p}{\partial T} + \mu\frac{\partial p}{\partial\mu}$, so that 
$$
  \frac{\partial p}{\partial T} \left( \frac{dT}{dr} + T \frac{d\P}{dr} \right) +
  \frac{\partial p}{\partial \mu} \left( \frac{d\m}{dr} + \mu \frac{d\P}{dr} \right) = 0\,.
$$
As $\partial p/\partial T$ and $\partial p/\partial\mu$ are independent functions, this is solved by 
\begin{equation}
\label{eq:Tmu-equil}
  T(r) = \bar T e^{-\Phi(r)}\,,\quad \mu(r) = \bar\mu \, e^{-\Phi(r)}\,,
\end{equation}
where $\bar T$ and $\bar\mu$ are constants of integration. This is a particular manifestation of the general statement that in equilibrium both $T$ and $\mu$ must be proportional to $1/\sqrt{g_{00}}$~\cite{RevModPhys.21.531}. Thus both $T$ and $\mu$ are expressed in terms of $\Phi$, and we in fact have only two unknown functions $m(r)$ and $\Phi(r)$, for which we have the two equations \eqref{eq:eett} and \eqref{eq:eerr}. Alternatively, eq.~(\ref{eq:Tmu-equil}) implies that $\tau\equiv T/\mu$ is a constant that does not depend on $r$. Then, for a given equation of state $p(T,\mu) = p(\tau \mu, \mu)$, we have
\begin{subequations}%
\label{eq:TOV}
\begin{align}
  & \frac{dm}{dr} = 4\pi r^2 \left(-p+T \frac{\partial p}{\partial T} + \mu \frac{\partial p}{\partial\mu} \right),\\
  & \frac{d\m}{dr} = -\mu \frac{\GN m}{r^2} \frac{1+ 4\pi r^3 p/m}{1- 2\GN m/r}.
\end{align}
\end{subequations}
At the origin $r = 0$ we set $m(0)=0$, and $\mu(0)=\mu_0$ to some constant which determines the pressure at the center. The equations~(\ref{eq:TOV}) are then integrated numerically starting from the origin. If there is a point $r=R$ where the pressure drops to zero, we have a star. The metric eq.\ (\ref{eq:ss2}) then describes the star for $r<R$, and has to be matched to the Schwarzschild metric at $R$, so that $m(R)=M$ is the total mass of the star. If the pressure never drops to zero, this means the equation of state does not allow for stars, and we have a spherical distribution of matter that fills the whole space. 

Another way to look at the TOV equations is to note that for the general spherically symmetric metric eq.\ \eqref{eq:ss1}, the Einstein tensor is schematically of the form $G_{\mu}^{\ \, \nu} = {\rm diag}(X, Y, Z, Z)$, which has to be matched to $T_\mu^{\ \, \nu} = {\rm diag}(-\epsilon, p, p, p)$. The TOV equations arise by enforcing $Y=Z$. On the other hand, for matter subject to a radial gravitational field one in general expects $T_{r}^{\ \, r} \neq T_\theta^{\ \,\theta}$. This is because the $r$-dependent metric makes the radial and the angular directions inequivalent, and there is no symmetry that would enforce $T_{r}^{\ \, r} = T_\theta^{\ \,\theta}$, or $Y=Z$. This difference between $T_{r}^{\ \, r}$ and $T_\theta^{\ \,\theta}$ which is not captured by the perfect fluid model of matter is precisely what is described by the effective Einstein's equations \eqref{eq:EE-1} in a systematic way.

Let us now incorporate the effects of the gravitational susceptibilities. Substituting the metric eq.\ \eqref{eq:ss2} into the effective Einstein's equations \eqref{eq:EE-1}, with the matter energy-momentum tensor given by eqs.\ \eqref{eq:Tmn}, \eqref{eq:cr-statics}, the equations for the potential and the mass function are
\begin{subequations}
\label{eq:spheq-22}
\begin{align}
&r^2 \big[ \e (f_1 - f_1') + p (f_2 - 2f_1')\big] + f_1 (2f_1' - f_2) \frac{2Gm}{r} +\big[ f_1(2f_1' + f_2) - 2(f_1'^{2} + f_1^2)\big] 2G \,\frac{dm}{dr} \nonumber\\
&+\big[f_1(4f_1' + 2f_1''-f_2-f_2') + f_1' (2f_1'' - 3f_2 + f_2') + f_2(f_2 - 2 f_1'')\big]r(r - 2Gm)\left(\frac{d\P}{dr}\right)^2 \nonumber\\
&+2(f_1'-f_1)(2f_1'-f_2)(r-2Gm) \, \frac{d\P}{dr}  = 0\, , \label{modEE1}\\
&r^3 p + 4f_1 Gm + (r-2Gm)\bigg[ 4(f_1' - f_1) + (2f_1' - f_2) r \, \frac{d\P}{dr}\bigg] r \, \frac{d\P}{dr} = 0\, . \label{modEE2}
\end{align}
\end{subequations}
Here $f_1 = 1/16\pi G + f_{\rm 1,m}(T,\mu)$, where $f_{\rm 1,m}(T,\mu)$ is the matter contribution. The susceptibility $f_3(T,\mu)$ does not contribute for a non-rotating configuration. The primes are defined by eq.~\eqref{defprime}, as before.
It is easy to see that on taking the zeroth order values for the susceptibilities, $f_1 = 1/16\pi G, f_2  = 0$ in eqs.\ \eqref{modEE1}, \eqref{modEE2} above, one recovers the standard TOV equations \eqref{eq:eett}, \eqref{eq:eerr}. The temperature and the chemical potential are still given by eq.~\eqref{eq:Tmu-equil}, thanks to the general definition \eqref{eq:uTmu}. Thus eqs.~\eqref{eq:spheq-22} are the leading-order modification of the TOV equations by the matter's response to curvature.

\section{Free fermions}

For a given microscopic theory, the susceptibilities $f_{1,2,3}(T,\mu)$ that appear in the effective Einstein's equations may be computed from equilibrium two-point functions of the energy-momentum tensor, as derived for example in~\cite{Kovtun:2018dvd}. In a non-interacting theory, the energy-momentum tensor is quadratic in the fundamental fields, and the susceptibilities are given by one-loop diagrams that can be readily evaluated in the Matsubara formalism~\cite{Romatschke:2009ng, Moore:2010bu, Moore:2012tc, Megias:2014mba, Buzzegoli:2017cqy, Kovtun:2018dvd, Shukla:2019shf}. 

The loop integrals will contain ultraviolet divergences, reflecting the need to renormalize the susceptibilities. Regulating the ultraviolet divergences with a large-momentum cutoff, we have $f_{n} = f_{n,{\rm UV}} + f_{n,{\rm m}}(T,\mu)$, where the matter contributions $f_{n,{\rm m}}(T,\mu)$ do not depend on the cutoff and vanish as $T,\mu\to0$. The cutoff-dependent contributions $f_{n,{\rm UV}}$ renormalize the parameters of the effective action. Clearly, as $f_1$ multiplies the Ricci scalar in the effective action, $f_{1,{\rm UV}}$ renormalizes the vacuum Newton's constant, so that $f_1=1/2\kappa^2 + f_{\rm 1,m}(T,\mu)$. The effective Newton's constant that appears in the effective Einstein's equations (\ref{eq:EE-1}) is then determined by 
\begin{equation}
\label{eq:km}
  \frac{1}{2\kappa_{\rm eff}^2} = \frac{1}{2\kappa^2} + f_{1,{\rm m}}(T,\mu)\,.
\end{equation}
The ultraviolet divergences in $f_2$, $f_3$, on the other hand, have no physical meaning in a Lorentz-invariant microscopic theory.%
\footnote{
  For Lorentz-violating matter that breaks the Lorentz invariance through a preferred timelike vector, $f_{2,{\rm UV}}$, $f_{3,{\rm UV}}$ would renormalize the corresponding coefficients in the gravitational effective action. Tuning the Lorentz breaking parameter to zero gets rid of the coefficients.
}
As the presence of the terms $a^2$, $\Omega^2$ in the effective action is thermal state-specific, so must be the susceptibilities $f_2$, $f_3$. Thus in the gravitational effective action we have $f_2 = f_{\rm 2,m}$, $f_3 = f_{\rm 3,m}$.

\begin{figure*}[t]
\includegraphics[scale=0.425]{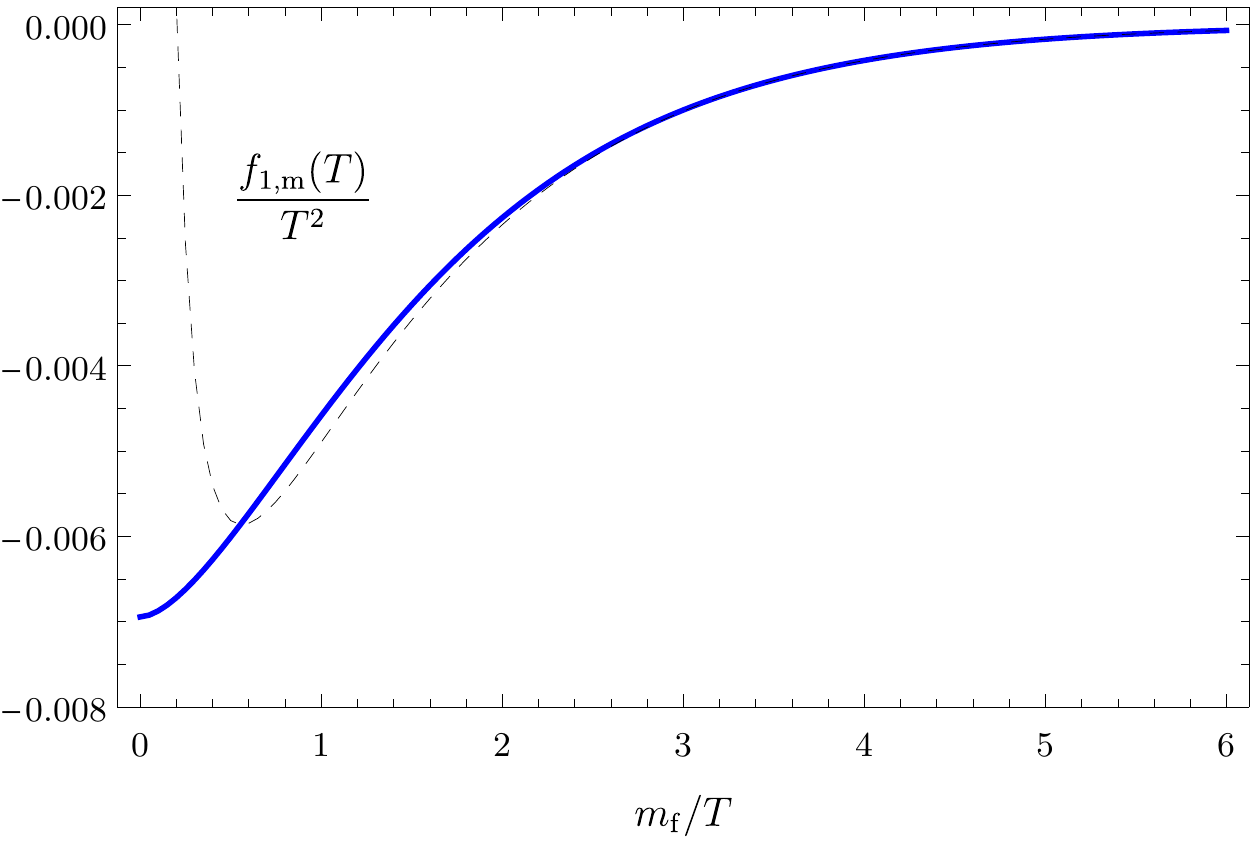}
\includegraphics[scale=0.425]{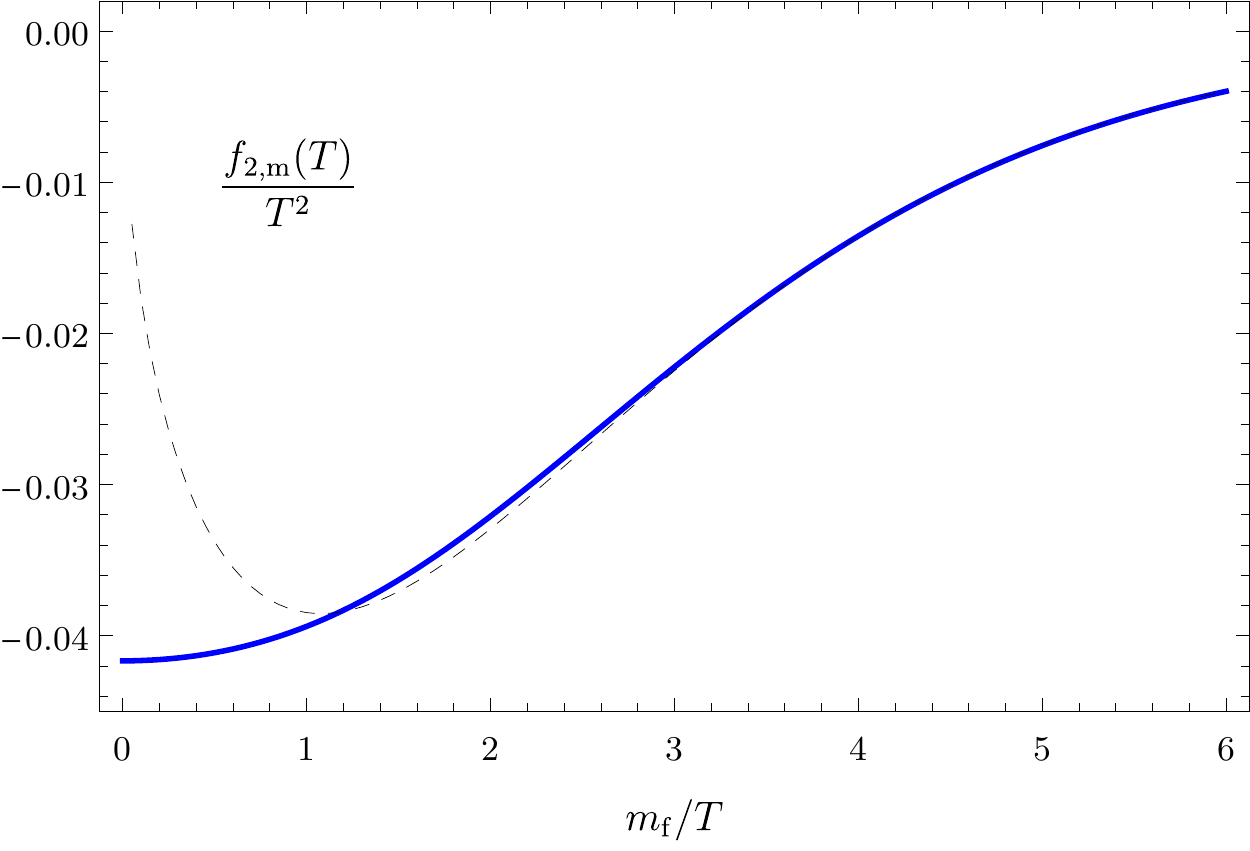}
\includegraphics[scale=0.425]{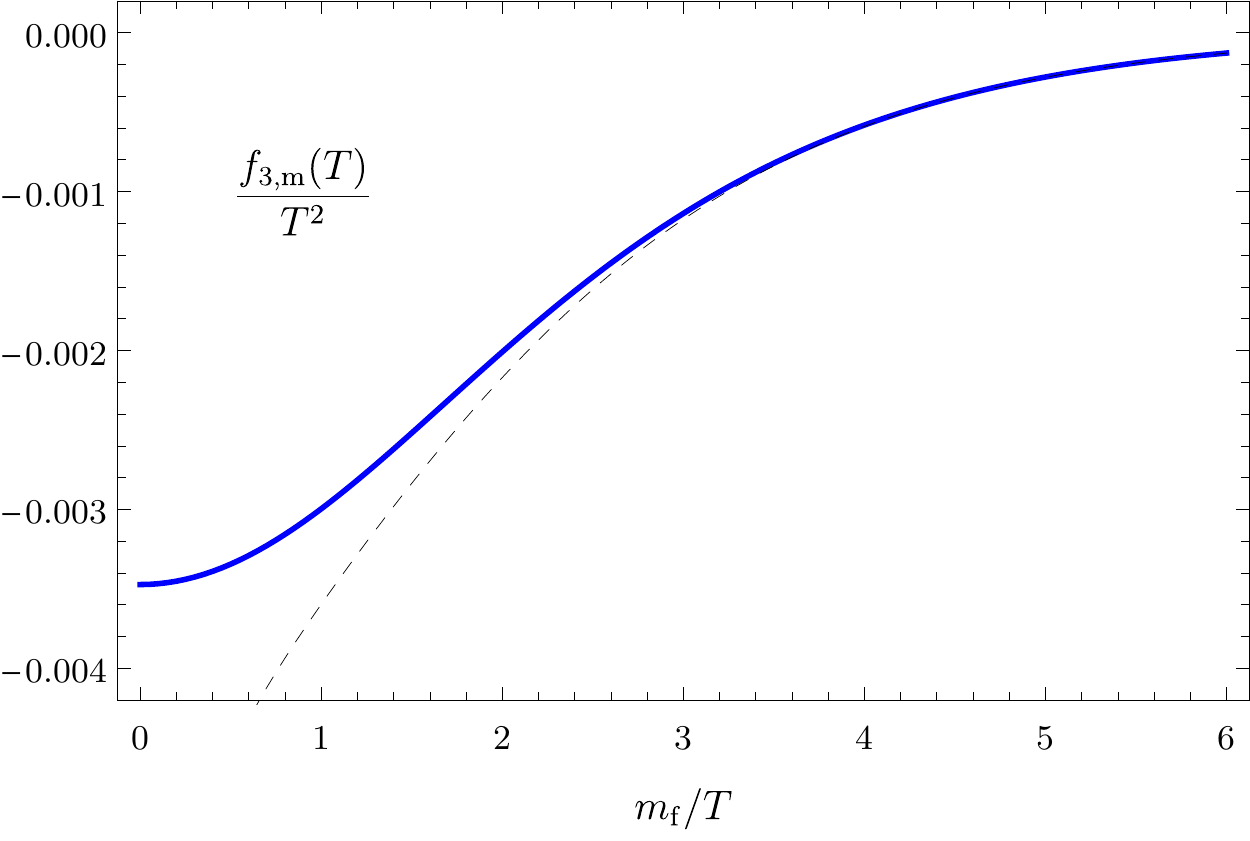}
\caption{
\label{fig:f123T}
The thermodynamic susceptibilities $f_{n,{\rm m}}(T)/T^2$ for a gas of free Dirac fermions of mass $\mf$ at $\mu=0$. The blues lines are the exact values from the one-loop diagrams, the dashed lines are the low-temperature approximations of eqs.~(\ref{eq:h1smallT}), (\ref{eq:h2smallT}), (\ref{eq:h3smallT}).
}
\end{figure*}

For concreteness, let us consider free Dirac fermions of mass $\mf$ in 3+1 dimensions. 
At $T=0$, the susceptibilities $f_n$ are only non-zero for $|\mu|>\mf$, just like the pressure $p(\mu)$. Evaluating the one-loop diagrams, one finds~(see e.g.\ \cite{Shukla:2019shf})
\begin{align}
 & f_{\rm 1,m} = -\frac{\mf^2}{48\pi^2} \left[\frac{|\mu|}{\mf^2} \sqrt{\mu^2 {-} \mf^2} - \ln \!\left(\frac{|\mu| + \sqrt{\mu^2 {-} \mf^2}}{\mf}\right) \right]\!, \\
 & f_{\rm 2,m} = \frac{|\mu|}{24\pi^2}\frac{(2\mf^2 - 3\mu^2)}{\sqrt{\mu^2 - \mf^2}} \,, \\
 & f_{\rm 3,m} = -\frac{|\mu|}{96\pi^2}\sqrt{\mu^2 - \mf^2} \,.
\end{align}
At $\mu=0$, we have $f_{n,{\rm m}}(T) = T^2 h_n(\mf/T)$ with dimensionless functions $h_n(x)$. Evaluating the one-loop diagrams, for $T\ll \mf$ one finds asymptotic expansions
\begin{align}
\label{eq:h1smallT}
 & h_1(x) \sim -\frac{e^{-x}}{4\sqrt{2}} \left( \frac{x^{1/2}}{3} + \frac{x^{-1/2}}{8} - \frac{5x^{-3/2}}{128} + \dots \right)\!,\\
\label{eq:h2smallT}
 & h_2(x) \sim -\frac{e^{-x}}{4\sqrt{2}} \left( \frac{x^{5/2}}{3} + \frac{9 x^{3/2}}{8} + \frac{235x^{1/2}}{128} + \dots \right)\!, \\
\label{eq:h3smallT}
 & h_3(x) \sim -\frac{e^{-x}}{16\sqrt{2}} \left( \frac{x^{3/2}}{3} + \frac{5 x^{1/2}}{8}  + \frac{35 x^{-1/2}}{128}\dots \right).
\end{align}
At high temperatures, $h_1(0) = -1/144$, $h_2(0) = -1/24$, $h_3(0)=-1/288$ \cite{Kovtun:2018dvd}. The functions $h_n(x)$ are shown in Fig.~\ref{fig:f123T}.

\begin{figure*}[t]
\includegraphics[height=3.85cm]{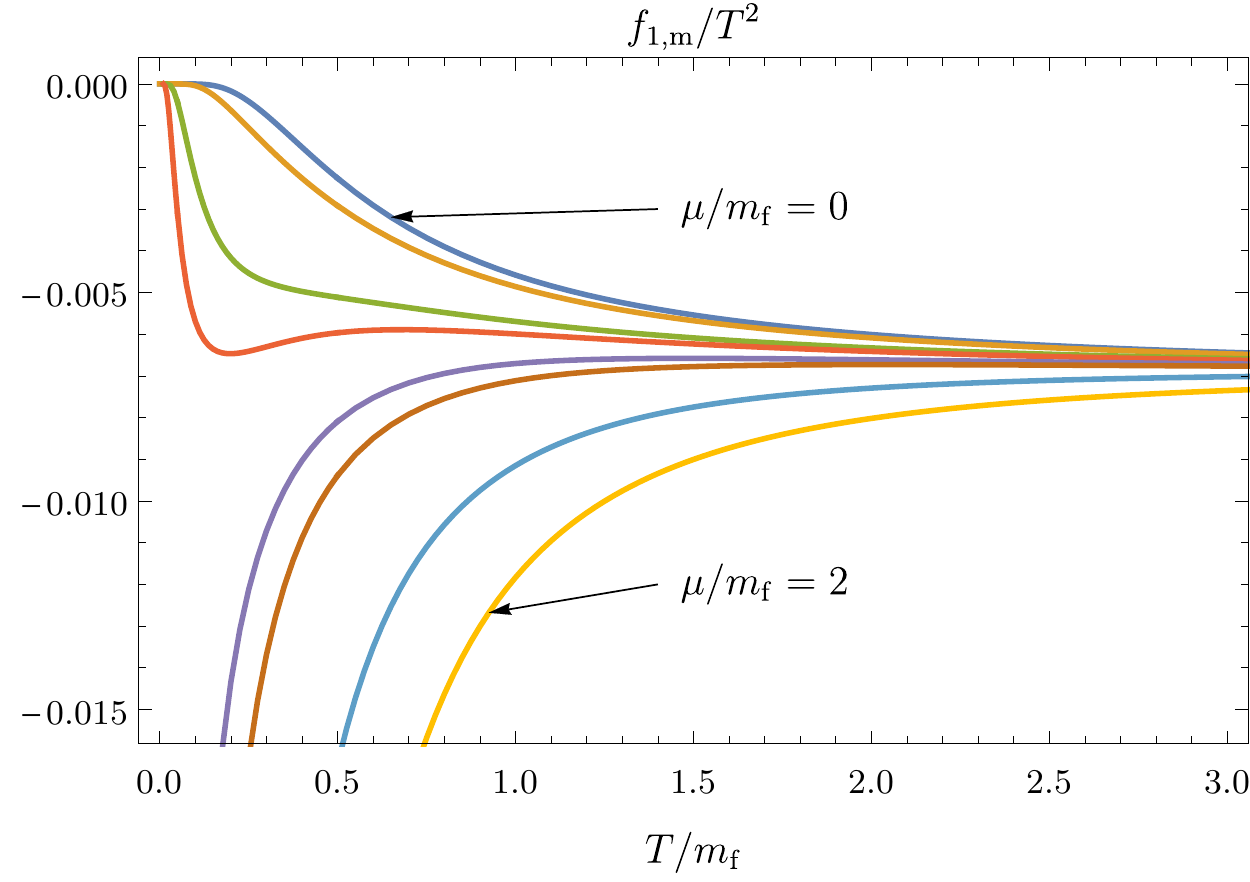}
\includegraphics[height=3.85cm]{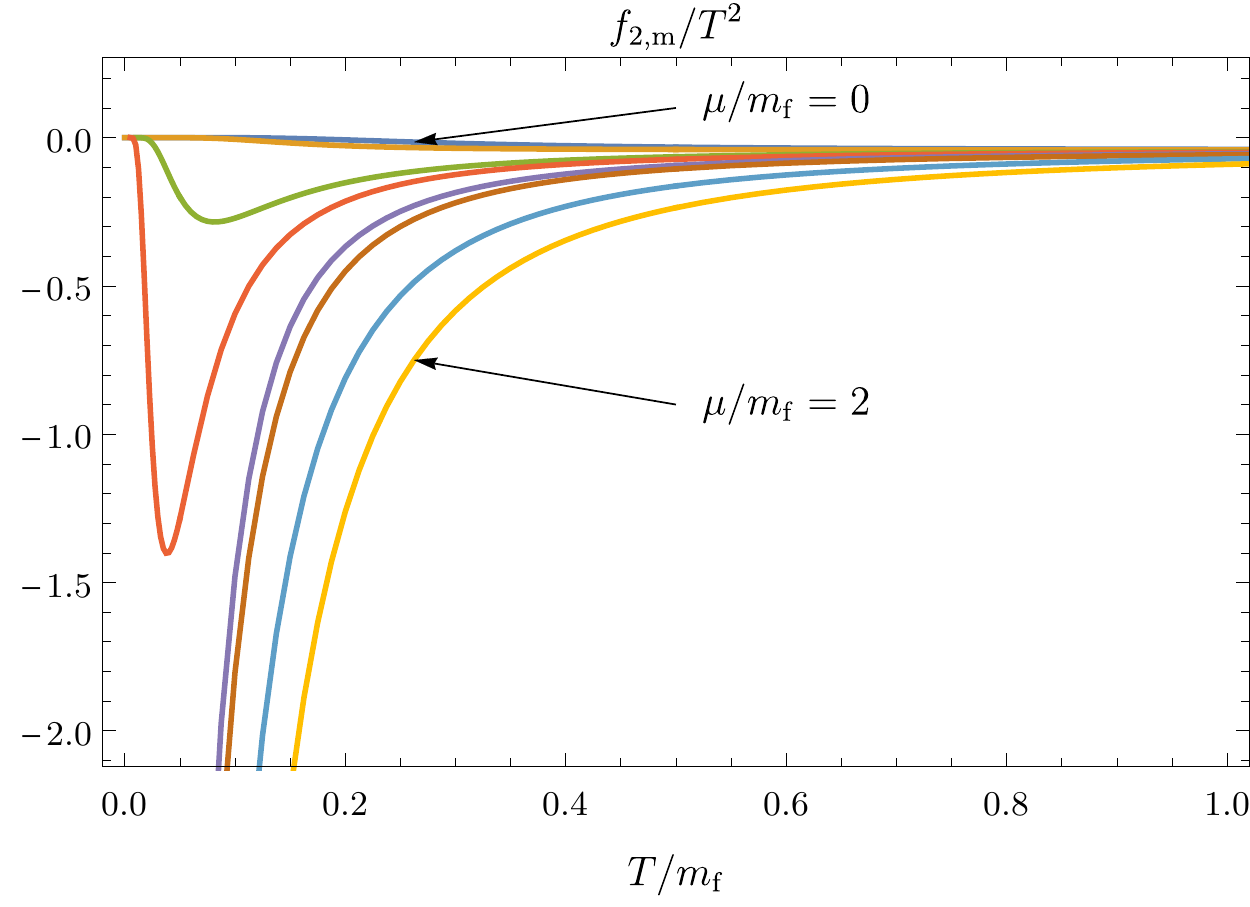}
\includegraphics[height=3.85cm]{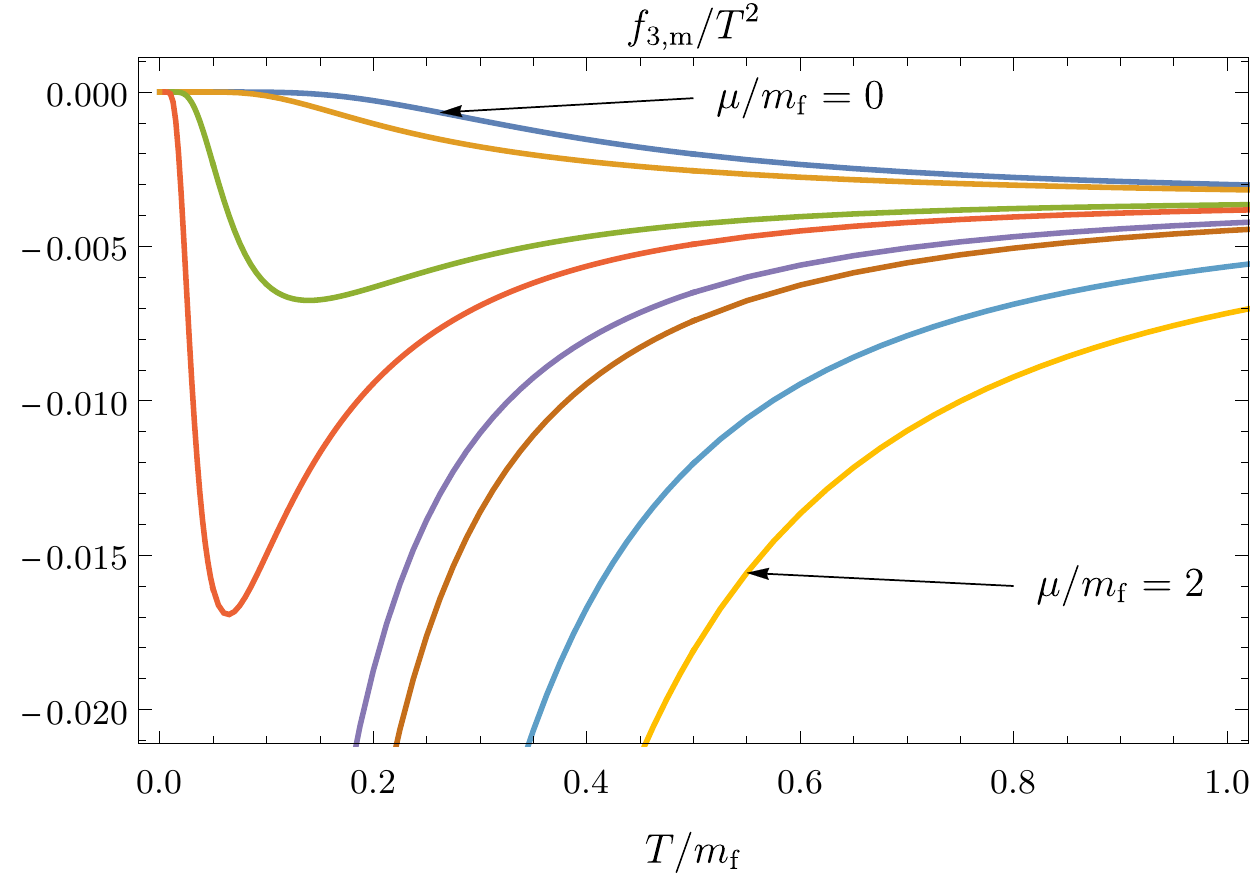}
\caption{
\label{fig:f123Tmu}
Thermodynamic susceptibilities $f_{n,{\rm m}}(T,\mu)/T^2$ for a gas of free Dirac fermions of mass $\mf$, at various values of $\mu/\mf$. The curves of different color correspond to the values $\mu/\mf = 0, 0.4, 0.8, 0.9, 1.1, 1.2, 1.6, 2.0$, from top to bottom in each plot. As $T\to0$, all curves go to zero for $\mu<\mf$, and diverge for $\mu>\mf$. 
}
\end{figure*}

At the leading order in the non-relativistic semiclassical limit $\frac{\mf}{T}\gg \frac{\mf-\mu}{T}\gg1$, the susceptibilities are proportional to the particle number density $n = (\mf T/2\pi)^{3/2}e^{(\mu-\mf)/T}$. Restoring the factors of $\hbar$, we have
\begin{align}
   f_{\rm 1,m} \sim -\frac{\hbar^2}{12} \frac{n}{\mf}\,,\ \ \ \ \
   f_{\rm 2,m} \sim -\frac{\hbar^2}{12} \frac{n\, \mf}{T^2}\,,\ \ \ \ \
   f_{\rm 3,m} \sim -\frac{\hbar^2}{48} \frac{n}{T}\,.
\end{align}
At non-zero $T$ and $\mu$, the susceptibilities $f_{n,{\rm m}}$ are plotted in Fig.~\ref{fig:f123Tmu}.

\section{Matter in anti-de Sitter space}
Consider uncharged conformal matter at temperature $T$ in four dimensions. The pressure is $p_{\rm m}(T) = c_0 T^4$, while the gravitational susceptibilities are $f_{\rm 1,m}(T) = \frac16 f_{\rm 2,m}(T) = \varphi_0 T^2$, where the constants $c_0$ and $\varphi_0$ are dimensionless in the natural units $\hbar=c=1$. For the conformally coupled real massless scalar field $c_0 = \frac{\pi^2}{90}$, and $\varphi_0 = 0$.%
\footnote{
For a massless scalar field in 3+1 dimensions, $\varphi_0$ is proportional to $(1-6\xi)$, where $\xi$ is the coupling to curvature, $\xi=1/6$ for the conformal coupling~\cite{Kovtun:2018dvd}.
}
For free massless Dirac fermions $c_0 = \frac78 \frac{4\pi^2}{90}$, and $\varphi_0 = -\frac{1}{144}$. For the black-body radiation (gas of non-interacting photons) $c_0 = \frac{2\pi^2}{90}$, and $\varphi_0 = -\frac{1}{36}$. See Ref.~\cite{Kovtun:2018dvd} for a summary of the calculations of $f_{\rm 1,m}$ and $f_{\rm 2,m}$ in free theories. For ${\cal N}=4$ supersymmetric $SU(N_c)$ Yang-Mills theory at infinite coupling and large $N_c$, a holographic calculation~\cite{Baier:2007ix} gives $c_0 = \frac{\pi^2}{8} N_c^2$, $\varphi_0 = -\frac{1}{16}N_c^2$. While $c_0$ must be positive for positive pressure, we are not aware of a general argument that would fix the sign of $\varphi_0$.

Let us now study the effects of conformal matter on the anti-de Sitter space AdS$_4$, with the cosmological constant $\Lambda = -3/\ell^2<0$. Einstein's equations in matter are applicable when 
\begin{align}
\label{eq:lll}
  l_{\rm Pl} \ll \lambda_{\rm th} \lesssim \ell\,,
\end{align}
where $l_{\rm Pl} = 1/M_{\rm Pl}$ is the Planck length, and $\lambda_{\rm th} = 1/T$ is the thermal wavelength. The first inequality is the applicability of classical gravity, and the second is the applicability of the derivative expansion for the matter's generating functional. Using $\alpha = \frac{c_0 T^4}{M_{\rm Pl}^2/\ell^2}$, and $\gamma = \varphi_0 T^2/M_{\rm Pl}^2$, Einstein's equations in matter are applicable when $|\gamma| \ll 1$ and $|\gamma| \lesssim \alpha$. We will see below that violating the inequalities (\ref{eq:lll}) will lead to quantitative criteria of the breakdown of hydrostatics, which are invisible for perfect-fluid matter.

The parameter $\alpha$ can be thought of as the ratio of the energy density in matter to the energy density in the cosmological constant. Thus the configuration is ``matter-dominated'' for $\alpha\gg1$ and ``cosmological constant-dominated'' for $\alpha\ll1$. A matter-dominated configuration implies $(T\ell)^2\gg M_{\rm Pl}^2/T^2 \gg 1$, and thus the effects of the gravitational susceptibilities are small. In fact, for $T\ell \gg 1$ the configuration would be on the high-temperature (black hole) side of the Hawking-Page phase transition~\cite{Hawking:1982dh}.  On the other hand, a configuration that is cosmological constant-dominated with $\alpha = c_0 (T\ell)^2 (T/M_{\rm Pl})^2 \ll 1$ can have $T\ell \sim 1$, and thus the effect of the gravitational susceptibilities can be noticeable. Note that the ``thermal AdS'' phase of the Hawking-Page transition exists for $T\ell\lesssim \frac{1}{\pi}$, while the perfect-fluid approximation in AdS is valid for $T\ell\gg 1$. Hence the low-temperature phase of the Hawking-Page transition can not be described by a perfect fluid in AdS. Perfect radiation fluids in AdS$_4$ were studied in Ref.~\cite{Page:1985em}. Einstein's equations in matter (\ref{eq:EE-1}) allow for a systematic calculation of the $1/(\ell T)^2$ corrections to the perfect-fluid approximation.

The static metric takes the form (\ref{eq:ss1}), and the equilibrium temperature is $T(r)=T_0/\sqrt{A(r)}$. Using $\rho\equiv r/\ell$, effective Einstein's equations (\ref{eq:EE-1}) give two equations for $A(\rho)$ and $B(\rho)$:
\begin{align}
\label{eq:ads-1}
  & \rho A' + A - AB(1{+}3\rho^2) - \alpha_0  \frac{\rho^2 B}{A} + \gamma_0 \left( \frac{(\rho A' - 2A)^2}{2A^2} - 2B \right) = 0\,, \\
\label{eq:ads-2}
  & \rho B' - B + B^2 (1{+}3\rho^2) - \alpha_0 \frac{3\rho^2 B^2}{A^2} \nonumber\\ 
  & + \frac{\gamma_0}{A^2} \left( \frac{5\rho^2 A'^2 B}{2A} -2A(\rho B' {+} B^2 {-} B) -4\rho BA' + \rho^2 A' B' -2\rho^2 A'' B \right) = 0 \,,
\end{align}
where $\alpha_0 \equiv c_0 (T_0\ell)^2 (T_0/M_{\rm Pl})^2$, and $\gamma_0 \equiv \varphi_0 T_0^2/M_{\rm Pl}^2$. Note that Eq.~(\ref{eq:ads-1}) is an algebraic equation for $B(r)$, which can be used to derive a second-order differential equation for $A(r)$. Without matter we have $\alpha_0 =\gamma_0 =0$, and Eqs.~(\ref{eq:ads-1}), (\ref{eq:ads-2}) are immediately solved to give 
\begin{align}
  B(r) = \frac{1}{1 - \frac{a'}{r} + \frac{r^2}{\ell^2}}\,,\ \ \ \ 
  A(r) = a \left( 1 - \frac{a'}{r} + \frac{r^2}{\ell^2} \right) \,,
\end{align}
where the integration constant $a'$ determines the mass of the AdS-Schwarzschild black hole. We will set $a'=0$ and consider matter at temperature $T$ in AdS, without the black hole. The integration constant $a$ can be absorbed into a rescaling of time. 

When $\gamma_0 =0$ and $\alpha_0 \neq0$, Eqs.~(\ref{eq:ads-1}), (\ref{eq:ads-2}) describe the backreaction of conformal matter, approximated as a perfect fluid, on AdS. The solution for $A(r)$ which is regular as $r\to0$ has the following expansions
\begin{align}
  & A(r\to0) = a + \left( a + \frac{\alpha_0}{a}\right) \frac{r^2}{\ell^2} + \frac{\alpha_0 (a^2 + \alpha_0)}{5a^3} \frac{r^4}{\ell^4} + \dots\,,\\
\label{eq:Ainf-1}
  & A(r\to\infty) = b_0 \frac{r^2}{\ell^2} + b_0 - b_1\frac{\ell}{r} + \dots\,,
\end{align}
with integration constants $a$, $b_0$, $b_1$. The constant $a$ determines the temperature at the center, $T(r=0) = T_0/\sqrt{a}$. The constant $b_0$ can be set to one by a normalization of time. The remaining constant $b_1$ determines the mass of the configuration. The full solution for $A(r)$ can be constructed numerically to determine the large-$r$ asymptotics of the metric and thus the mass of the configuration for different $\alpha_0$~\cite{Page:1985em}. For a solution that is regular at the center, the mass of the matter configuration (in units of $\ell/(2G) \sim M_{\rm Pl}^2 \ell$) is $m(a) = b_1/b_0$. 

With both $\gamma_0$ and $\alpha_0$ nonzero, Eqs.~(\ref{eq:ads-1}), (\ref{eq:ads-2}) describe the backreaction of conformal matter, taking into account the leading $1/(T\ell)^2$ correction to the perfect fluid model of equilibrium matter. The solution for $A(r)$ which is regular as $r\to0$ has the following expansions
\begin{align}
\label{eq:A0-2}
   & A(r\to 0) = a + \frac{a^2 {+} \alpha_0 {-} 4\gamma_0 a}{a {-} 6\gamma_0} \frac{r^2}{\ell^2} +  \frac{(\alpha_0 {+} 2 a \gamma_0) (a^2 {+} \alpha_0 {-} 4a \gamma_0) (a{-}18\gamma_0) }{5a(a {-} 6\gamma_0)^3} \frac{r^4}{\ell^4} + \dots\,,\\
\label{eq:Ainf-2}
  & A(r\to\infty) = b_0 \frac{r^2}{\ell^2} + b_0 - b_1\frac{\ell}{r} + \dots\,.
\end{align}
The leading-order large-$r$ asymptotics are not modified compared to the perfect-fluid case, and the difference between Eqs.~(\ref{eq:Ainf-1}) and (\ref{eq:Ainf-2}) only comes at order $O(\ell^2/r^2)$ and higher.  The equation for $A(r)$ can then be integrated numerically starting with the initial condition (\ref{eq:A0-2}) in order to obtain the mass $m(a) = b_1/b_0$. 

The constants $a$ and $\alpha_0$ must be positive, while the sign of $\gamma_0$ is not a priori determined. The third term in the expansion (\ref{eq:A0-2}) vanishes for $\gamma_0 = -\alpha_0/(2a)$. In fact, for this value of $\gamma_0$, all the terms in the expansion (\ref{eq:A0-2}) except for the first two vanish identically, and one can easily check that the AdS space $A(\rho) = a(1 +\rho^2)$, $B(\rho) = (1+ \rho^2)^{-1}$ is an exact solution to Eqs.~(\ref{eq:ads-1}), (\ref{eq:ads-2}).%
\footnote{
Similarly, when $a^2 + \alpha_0 = 4\gamma_0 a$, one finds an exact solution $A(\rho) = a$, $B(\rho) = (1+2\rho^2)^{-1}$, which is not asymptotically AdS. This can happen for $\gamma_0^2>\alpha_0/4$, and at sufficiently high central temperature, $a<4\gamma_0$.
}

\begin{figure}
\includegraphics[width=0.48\textwidth]{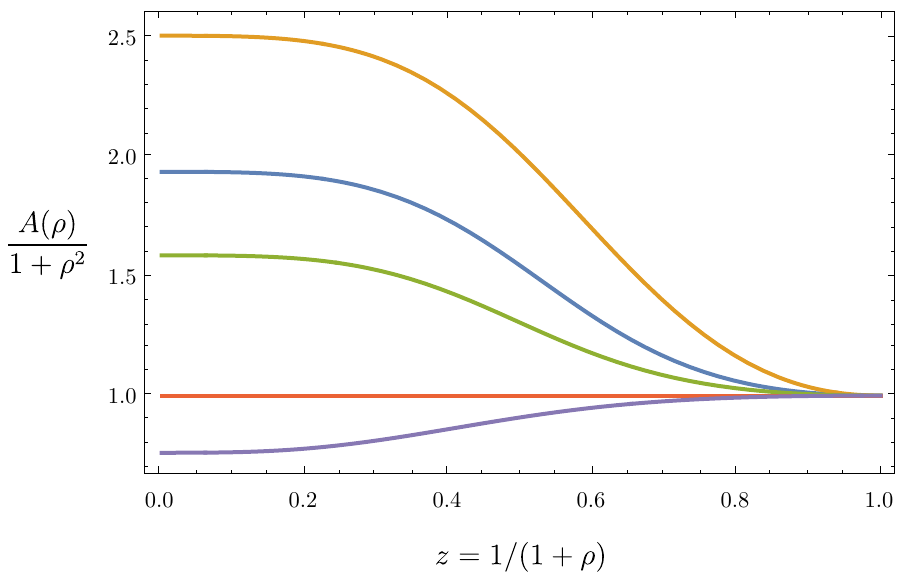}
\hspace{0.02\textwidth}
\includegraphics[width=0.48\textwidth]{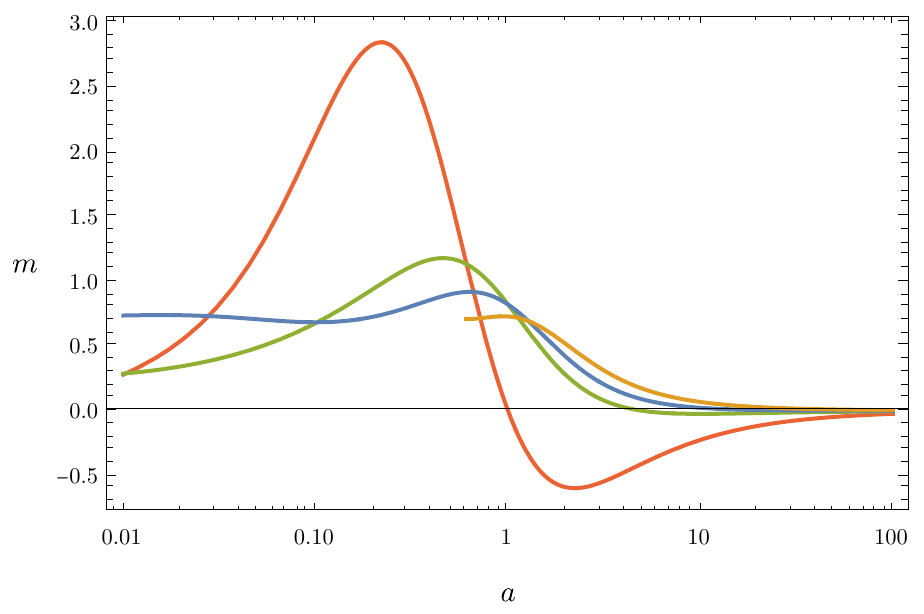}
\caption{
\label{fig:ads}
Left: The metric component $A(r)=-g_{00}(r)$ as a function of $z \equiv 1/(1+r/\ell)$ for $a=1$ and $\alpha_0=1$, obtained by solving Eqs.~(\ref{eq:ads-1}), (\ref{eq:ads-2}) numerically. From top to bottom, the curves correspond to $\gamma_0 = 0.1, 0, -0.1, -0.5, -1$. The mass is determined by the behaviour of the curve at small $z$. As the value of $\gamma_0$ falls below $-\alpha_0/(2a)$, the curve changes from concave to convex, and the mass turns negative. Right: the mass of the matter configuration at $\alpha_0{=}1$ as a function of the constant $a$ which determines the central temperature, $T(r{=}0)=T_0/\sqrt{a}$. From top to bottom at large $a$, the curves correspond to $\gamma_0 = 0.1, 0, -0.1, -0.5$. For the perfect fluid ($\gamma_0 = 0$) the mass is bounded from above by a number of order one~\cite{Page:1985em}. For $0<a<6\gamma_0$, the space is no longer asymptotically AdS, and the curve is cut off.
}
\end{figure}

The total mass is positive for $\gamma_0 > -\alpha_0/(2a)$, and negative for $\gamma_0 < -\alpha_0/(2a)$, as illustrated in Fig.~\ref{fig:ads}. Correspondingly, the energy density of matter $T^{00}_{\rm m}$ in the right-hand side of Eq.~(\ref{eq:EE-1}) turns from being positive everywhere in space to being negative everywhere in space, as $\gamma_0$ crosses the critical value $-\alpha_0/(2a)$. Thus we see that for matter with negative $\varphi_0$ (such as free photons), the mass of the static configuration of matter becomes negative for
\begin{equation}
\label{eq:Tl}
  T(r{=}0) < \frac{1}{\ell} \left( \frac{2|\varphi_0|}{c_0}\right)^{1/2} \,.
\end{equation}
Of course, one does expect on physical grounds that the macroscopic fluid description of matter should become physically inadequate when the temperature is sufficiently small, $T\ell \lesssim 1$. However, one has to go beyond the perfect-fluid approximation of matter in order to see a quantitative criterion for the breakdown of hydrostatics, Eq.~(\ref{eq:Tl}). This breakdown corresponds to the violation of the second inequality in Eq.~(\ref{eq:lll}). 

Another point to note is that the small-$r$ expansion (\ref{eq:A0-2}) is really an expansion in $(r/\ell)/(a{-}6\gamma_0)$, and therefore its radius of convergence shrinks to zero as $\gamma_0$ approaches $a/6$ from below. For $\gamma_0 > a/6$, the equation for $A(\rho)$ can still be integrated numerically by using the expansion (\ref{eq:A0-2}) and starting at a sufficiently small value of $\rho$. The result appears to be a solution which has $b_0 = 0$, and therefore the space that is not asymptotically AdS. This happens for 
\begin{equation}
\label{eq:TMpl}
  T(r{=}0) > \frac{M_{\rm Pl}}{(6\varphi_0)^{1/2}}\,,
\end{equation}
which we interpret as the breakdown of the classical description at high temperatures. The criterion (\ref{eq:TMpl}) arises for matter with $\varphi_0>0$, and corresponds to the violation of the first inequality in Eq.~(\ref{eq:lll}).

\section{Concluding Comments}
Starting with the equilibrium generating functional, we have set up a framework to describe classical gravitational fields with matter present. Equations (\ref{eq:EE-1}) with the energy-momentum tensor (\ref{eq:Tmn}), (\ref{eq:cr-statics}) are the gravitational analogues of the electro- and magneto-statics in matter. The temperature- and density-dependent correction to the Newton's constant is given by eq.~(\ref{eq:km}), with $f_{\rm 1,m}$ for Dirac fermions plotted in Fig.~\ref{fig:f123Tmu}, leftmost panel. 

For normal quantum matter in flat space the effects of the gravitational susceptibilities will be small. Consider the example of the TOV equations. For the perfect fluid approximation, the Einstein's equations are schematically of the form $1/L^2 \simeq m^4/M_{\rm Pl}^2$ where $L$ is a length scale, with $L^2$ coming from the two derivatives in the Einstein tensor, $1/M_{\rm Pl}^2 = 8\pi G$, and $m$ is the energy scale of the matter (this gives $L\simeq M_{\rm Pl}/m^2$, an estimate of a neutron star size for $m\simeq 1 {\rm GeV}$). With the gravitational susceptibilities taken into account, schematically
$$
  \frac{1}{L^2} \simeq \frac{1}{M_{\rm Pl}^2} \left( m^4 + \frac{m^2}{L^2}\right)\,,
$$
and the corresponding effects are $(m/M_{\rm Pl})^2$ suppressed. 
This is expected, as the gravitational susceptibilities $f_n$ are quantum-mechanical in nature -- for classical particles in external gravitational field the equilibrium partition function does not depend on the Riemann curvature. 

The implications are more interesting in Anti-de Sitter space, where we have studied hydrostatic configurations of uncharged conformal matter, such as a gas of free photons. The perfect fluid model of matter is only physically applicable for $T\ell \gg1$, where $\ell$ is the AdS radius. However the Einstein equations coupled to perfect-fluid matter exhibit no pathologies even for $T\ell \lesssim 1$, and one might be tempted to treat matter in AdS as a perfect fluid even at low temperatures. The gravitational susceptibilities studied in this paper allow for a systematic study of the $1/(T\ell)^2$ corrections to the perfect-fluid approximation. For free photons and fermions (as well as for the strongly coupled supersymmetric Yang-Mills theory), the effect of the gravitational susceptibilities is to reduce the total mass of the hydrostatic configuration, and eventually to drive the mass negative at small $T\ell$, see Eq.~(\ref{eq:Tl}). We interpret the negative mass as the breakdown of macroscopic hydrostatics. For example, for black-body radiation in AdS$_4$, macroscopic hydrostatics explicitly breaks down for 
\begin{equation}
  T(r{=}0) \ell < \frac{\sqrt{5/2}}{\pi}  \approx 0.5\,.
\end{equation}
For the strongly coupled supersymmetric Yang-Mills theory in AdS$_4$, hydrostatics breaks down for
\begin{equation}
  T(r{=}0) \ell < \frac{1}{\pi}\,.
\end{equation}
Curiously, the last expression formally coincides with the onset of the Hawking-Page transition. It is satisfying to see how basic thermal physics of gravitational susceptibilities ``restores justice'', and explicitly disallows for the macroscopic hydrostatics at small $T\ell$.

{\em Note added ---} After this paper was completed, we became aware of the preprint \cite{Baier:2019ciw} that uses a simplified version of the Einstein's equations in matter in a cosmological setting.

\acknowledgments
The work was supported in part by the NSERC of Canada.

\bibliography{EERefs}
\end{document}